\documentclass{article}
\usepackage{graphicx}

\newcommand{\no}{\noindent}

\newcommand{\beq}{\begin{equation}}
\newcommand{\eeq}{\end{equation}}
\newcommand{\beqn}{\begin{eqnarray}}
\newcommand{\eeqn}{\end{eqnarray}}
\newcommand{\beqns}{\begin{eqnarray*}}
\newcommand{\eeqns}{\end{eqnarray*}}
\newcommand{\Var}{\mbox{Var}}

\newcommand{\pa}{\emph{P. aeruginosa}}
\newcommand{\mt}{\emph{M. tuberculosis}}
\newcommand{\etal}{\emph{et al.}}

\begin{document}

\title{Estimating the Number of Essential Genes
in Random Transposon Mutagenesis Libraries}

\author{O. Will\footnote{Allan Wilson Centre, University of Canterbury, Christchurch, New Zealand. E-mail: owill4@yahoo.com} and M.~A. Jacobs\footnote{Department of Medicine, University of Washington Genome Center, Seattle, WA 98195 USA. E-mail: mikejac@u.washington.edu}}

\maketitle

\no \large{\textbf{Abstract}}

\no Biologists use random transposon
mutagenesis to construct knockout libraries for bacteria. Random
mutagenesis offers cost and efficiency benefits over the standard
site directed mutagenesis,\cite{jacobs:pa03} but
one can no longer ensure that all the nonessential genes will appear in the library.  In random libraries for
haploid organisms, there is always a class of genes for which
knockout clones have not been made, and the members of this class are either
essential or nonessential. One requires statistical methods to estimate the number of
essential genes. Two groups of
researchers, Blades and Broman\cite{blades:essOrf02} and Jacobs \etal,\cite{jacobs:pa03} independently and simultaneously
developed methods to do this. Blades and Broman used a Gibbs sampler and Jacobs \etal used a parametric bootstrap.
We compare the performance of these two methods and find that they both depend on having an accurate probabilistic model for
transposon insertion or on having a library with a large number of clones.
At this point, we do not have good enough probabilistic models so we must build libraries that have at least five clones per open reading frame to accurately estimate the number of essential genes.

\section{Introduction}
Scientists are creating knockout clonal
libraries for many micro-organisms.\cite{gerdes:ess03,giaever:sc02,hutchison:myco99,kobayashi:bs03} Usually, researchers follow a site directed
mutagenesis approach in which each clone had a predetermined open
reading frame (ORF) disrupted. Jacobs \etal\cite{jacobs:pa03} generated a library for
\emph{Pseudomonas areuginosa} cheaply and efficiently by random transposon mutagenesis. Rather than specifying which gene would be knocked out before mutation, they randomly mutated the genome and determined the location of the gene knockout afterwards.  

Each clone in a random mutagenesis library has a single transposon insertion. A transposon is a DNA sequence that can jump within chromosomes and between them. Biologists have built transposons with stop
codons that upon insertion in an ORF terminate protein translation.
When the insertion is random, one requires statistical methods to help construct the clonal library. Foremost, one would like to know how many clones to make. The number of clones made for the \pa library was based on the estimated number of essential genes.

Biologists and clinicians are interested in essential genes because they make good
candidates for drug targets.\cite{fraser:def04}  Essential genes never appear knocked out in random mutagenesis libraries because bacteria are haploid organisms. As a tautology, once an essential gene is knocked out, the bacterium cannot live. Besides essential genes, there are also nonessential genes that have yet to appear as knockouts in the library because of chance.
Jacobs \etal believed that a large number of clones should be created to get a good
estimate of the number of the essential genes. They made about
about five clones per ORFs and used a parametric bootstrap to estimate the number of essential genes. In a concurrent experiment, Lamichhane \etal\cite{lamichhane:mt03} used a Gibbs Sampler developed by
Blades and Broman\cite{blades:essOrf02} and claimed that they only
required 0.3 clones per ORF to obtain a good
estimate of number of essential genes in \emph{Mycobacterium
tuberculosis}. 

The \pa clinical isolate, PAO1, used to create the clonal library has 5,570 ORFs\cite{stover:pa00} and the \mt isolate has 4,250 ORFs. Prokaryotic genomes do not have introns so
stretches of DNA longer than 300 bp beginning with a methionine codon
and lacking a stop codon strongly indicate the existence of an
ORF. We assume that all the ORFs and their locations are known.

In this paper, we look at the fit of the probability model underpinning both estimation procedures and compare the accuracy and precision of the parametric bootstrap
to the Gibbs sampler. The \mt
library has 1,403 unique insertions, which corresponds to 1,839 insertions in the 
PAO1 library for the same coverage of $1839/5570=0.3\mbox{X}$. The PAO1 library
has 30,100 unique insertion locations; a coverage of $30100/5570 =
5.4\mbox{X}$ unique insertions per ORF. Both the parametric bootstrap and the Gibbs sampler
are based on the same multinomial insertion model. Considering
only unique positions, the multinomial model fits the PAO1 data at
the 0.3X coverage level. However, if we use all 30,100 insertions
to test model fit, we reject the hypothesis that the bin
probabilities used in the model are correct. The multinomial model
requires the ability to determine the probability that an
insertion lands in an individual ORF based solely on the
annotation (i.e. ORF length). We show that one can predict the
insertion probabilities but not accurately enough to estimate the
number of essential genes at coverage levels lower than 5.0X.

We analyze how effectively both methods estimate the
number of essential genes at lower than 5.4X coverage by using the empirical distribution of the
unique positions in the PAO1 library,  \emph{i.e.} drawing samples
without replacement from the list of insertion locations, in spite of the evidence against the multinomial insertion model. We find a
bias-variance trade-off in that the parametric bootstrap is more accurate
and the Gibbs sampler is more precise. At the 5.4X coverage level, the parametric bootstrap's 
estimate of the number of essential genes agrees with the estimate
from Gibbs sampler. Once
coverage is high enough, we are confident that one can
estimate the number of essential genes. The Gibbs sampler exhibits
bizarre behavior at coverage levels below 0.3X which we attribute
to mis-specification of the insertion probabilities.
Through simulation, we compute the necessary coverage for two
endpoints: Checking model accuracy and estimating the
number of essential genes.

\section{Empirical Verification of the Multinomial Model}
The data we use for this paper consist of the 42,240 mutants created for the PAO1 library. The locations of
36,154 of them have been mapped and 30,100 land in unique places. Of the
unique locations, 27,264 are inside ORFs and the other 2,836 are
between them. The inserts hit 4,895 of the 5,570 ORFs internally
so there are 675 candidate essential genes. We only consider unique locations, because there were
many more duplicate sites than expected by chance. We feel this is due to local contamination effects created by the high-throughput technology used to create the PAO1 library.

The exact
chemical mechanism for transposon insertion is not important for this
paper except that biologists putatively maintain that the elements
insert randomly and uniformly at specific targets in the genome.\cite{miller:mobile04} The elements used in the PAO1 library insert at any base pair and the element used in the \mt library inserts at TA dinucleotides. Since the PAO1 and \mt genomes are sequenced
and annotated,\cite{stover:pa00} we can locate the
exact position of insertion and assume that the location of the
transposable element can unambiguously be determined. Biologists engineer the transposons to have stop codons terminating the translation of
the protein. However, transposons that landed in the distal
portion of the ORF may have only stunted the protein allowing it
to function. Hence, there are varying definitions as to what
constitutes a knockout. Generally, we use the most liberal definition that
an insert landing anywhere within the ORF knocks out the protein function. Lamichhane \etal consider insertions that land in the
last 100 bp or 20\% of the ORF not to knockout the gene
(the ``$5'$80\%-$3'$100-bp'' rule). We also look at definitions in
which knockouts occur only when the insert landed in the middle
80\% or 60\% of the ORF.

A multinomial
distribution naturally approximates the biology of the transposable
elements. We let $k$ be the number of open reading frames in an
organism's genome, $n$ be the number of mutants assayed, and $m$
be the number of essential ORFs. The $m$ is unobservable and is
the focus of the estimation procedures. Based on the biology of the transposons used in the PAO1 library, we believe that the probability of hitting a
nonessential gene is proportional to its length in base pairs. One caveat that prevents the joint density of
number of times the ORFs are hit from being a true multinomial is
that the ORFs can overlap and an insertion in the overlap
knocks out both genes. Like Blades and Broman,\cite{blades:essOrf02} we let $\mathcal{G} = (g_1,\ldots,g_k)$ be
the vector of zeros and ones indicating whether an ORF is
nonessential with 0 indicating essentiality. We note that $\sum
g_i = k-m$ and define $X = (x_1,\ldots,x_k)$ to be the number of
insertions per ORF in regions that are not shared and $Y =
(y_1,\ldots,y_k)$ to be the number of insertions in regions shared
between gene $i$ and $i+1$. $y_k$ is the number of insertions into
the region shared by gene $k$ and the first gene,
because prokaryotic chromosomes are circular. We let
$(p_1,\ldots,p_k)$ be the probabilities that a transposable
element inserts into an ORF given that it is nonessential and let
$(q_1,\ldots,q_k)$ be the probability that it inserts into the
region shared by gene $i$ and $i+1$. Often it is convenient to
model the intergenic region as the $k^{\mbox{th}}$ ORF and then
the $q$'s have to be changed appropriately. We note that $\sum
x_i+y_i = n$ and $\sum p_i+q_i = 1$. Given $\mathcal{G}$, the
distribution of $(X,Y)$ is
\begin{equation}\label{dist}
P (X,Y|\mathcal{G}) = \frac{\prod_{1 \le i \le
k}(p_ig_i)^{x_i}(q_ig_ig_{i+1})^{y_i}}{(\sum_{1 \le j \le k}
p_jg_j+q_jg_jg_{j+1})^n}.
\end{equation}

\begin{figure}[ht]
%\begin{center}
\resizebox{4.7in}{2.9in}{\includegraphics{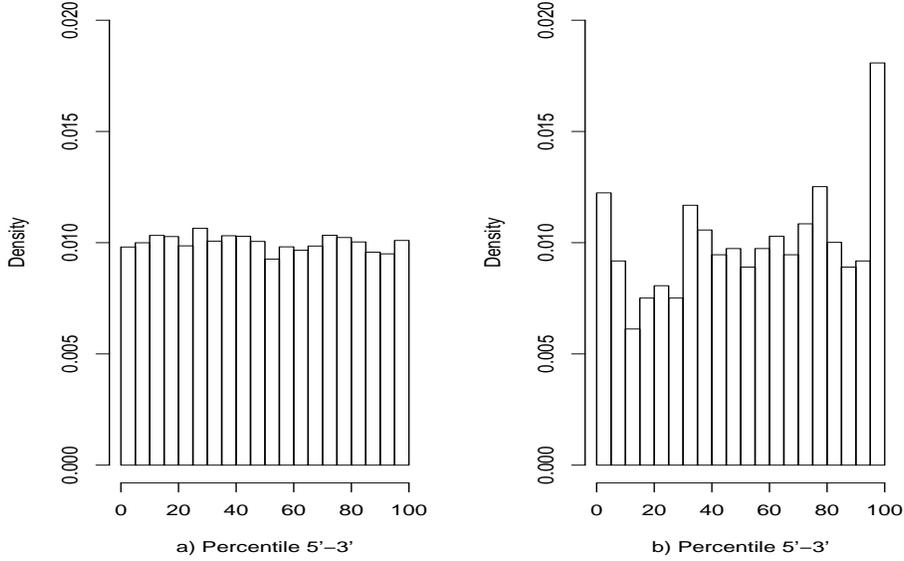}}
\caption{Hit distribution relative to the position within the ORF.
The histogram on the left (a) includes all insertions within an
ORF and the histogram on the right (b) includes insertions in ORFs
that are only hit once.}\label{histOrf}
%\end{center}
\end{figure}

We want to verify that the multinomial insertion model with a
class of unhitable essential genes fits the unique locations. We should check
that the unique locations occur uniformly and randomly in the
genome and that the probability of hitting an ORF can be computed
by dividing the length of the ORF by the number of bp in the
genome. First, 90.6\% of the unique locations are in coding
regions which is consistent with the 89.4\% of the genome that is
actually in coding regions.\cite{stover:pa00} We plot a histogram of the
location of the insertion sites within genes hit (Figure \ref{histOrf}).
The transposons appear to be inserting at each bp with equal
probability even though, according to the Kolomogorov-Smirnov
test, the empirical distribution is not consistent with the
Uniform distribution $(D=0.0089,\ p=0.026)$. Unfortunately, we
cannot use the intergenic region for testing the distribution of insertion locations since it is difficult to annotate.
The region is likely populated with regulatory elements that are
not well defined. 

We focus on the ORFs and define a modified goodness-of-fit statistic to accommodate the unobserved essential genes. We define $\chi^{2*}$ to be the goodness-of-fit statistic on only
those ORFs that are hit more than once. From large sample theory,
we know that if the number of insertions, $n$, is large then
\[ \sum_{1\le i \le k}\frac{(x_i-E_i)^2}{E_i} + \sum_{1\le i \le
k}\frac{(y_i-E_i)^2}{E_i}\sim \chi^2_{k'-1}. \]

$k'$ is number of ORFs and ORF overlaps that have one or more
targets. If one conditions on only a subset of the bins being
occupied in a multinomial, then the distribution of counts for
these bins are Multinomial $(n,p^*_1,\ldots,p^*_{k^*})$ with $p_i^*
= p_i/\left(p_1+\cdots+p_{k^*}\right)$. For example, if $k =
4$ and we condition on only bins 1 and 2 being occupied, then
$p^*_1 = \frac{p_1}{p_1+p_2}$ and $p^*_2=\frac{p_2}{p_1+p_2}$. If
we knew which genes were nonessential then we could test how well
the multinomial fits by looking at a $\chi^2$ on the conditioned
multinomial. Define \beq\label{chiStar} \chi^{2*} =
\sum\frac{(x^*_i-E^*_i)^2}{E^*_i}+\sum\frac{(y^*_i-E^*_i)^2}{E^*_i}
\eeq

\no where $x^*_i$ and $y^*_i$ are the counts of insertions in ORFs
or overlap regions that have been hit at least once. $E_i^* =
np_i^*$ in which $p_i^*$ is the recomputed probability of hitting
a region given that those regions with at least one insertion were
the only ones that could have been hit. We can show that the
asymptotic distribution of $\chi^{2*}$ is $\chi_{k'-m-1}$ by use of
one of the Slutsky Theorems.\cite{ferguson:large96} The
distribution is dependent upon the number of essential genes. Using the limiting distribution, we construct a very conservative
$\alpha$-level test for rejecting the null hypothesis that the
joint distribution of the number of times an ORF is hit is
Multinomial $(n,p^*_1,\ldots,p^*_{k^*})$. We reject the multinomial model if $\chi^{2*}
> \chi^2_{k'-1,1-\alpha}$, which occurs with probability less than
$\alpha$ if the null hypothesis is true.

If we compute the $\chi^{2*}$ goodness-of-fit
statistic  for ORFs hit anywhere internally at least once, we get a
value of 9663.9 on 6324 degrees of freedom. Under the null
hypothesis that that the multinomial insertion model is correct,
the 0.95 quantile for the $\chi^2_{6324}$ distribution is 6510.1,
and we safely reject the hypothesis that the insertions are
following a multinomial distribution whose probabilities are
computed from the length of the ORF divided by the number of base
pairs in the genome. Model-fitting with 30,100 observations is
often dangerous, but we use a conservative test
and found a value well within the critical region. We compute the
$\chi^{2*}$ goodness-of-fit statistic for other ORF definitions
(Table~\ref{chiFit}) and for each, we reject that the insertion probabilities based on gene length. For the rest of the paper, we use the most inclusive
definition of a knockout and count insertions anywhere within the
ORF.

\begin{table}
\caption{Goodness-of-Fit for Different Knockout Definitions \label{chiFit}}
\begin{center}
\begin{tabular}{ccrcc}
  \hline
  \hspace{1in} & Model & $\chi^{2*}$ Fit & DOF &  \hspace{1in} \\
  \hline
   \hspace{1in} & Entire ORF & 9663.9 & 6324 &  \hspace{1in} \\
   \hspace{1in} & Lamichhane & 7621.8  & 5570 &  \hspace{1in}  \\
  \hspace{1in}  & 5' 10\%-3' 10\% & 8487.5 & 5571 &  \hspace{1in} \\
   \hspace{1in}  & 5' 20\%-3' 20\% & 6733.1 & 5570 &  \hspace{1in} \\
 \hline
\end{tabular}
\end{center}
Different models, the modified $\chi^{2*}$ fit, and
number of ORFs added to the number of overlaps. Entire ORF counts
insertions that land anywhere within an ORF, Lamichhane counts
only those that do not land in the last 100 bp or the distal
20\%, 5' 10\%-3' 10\% counts those that land in the middle 80\%,
and 5' 20\%-3' 20\% counts those that land in the middle 60\%.
\end{table}

\begin{figure}[ht]

(a) \rotatebox{270}{\resizebox{2.8in}{4.0in}{{\includegraphics{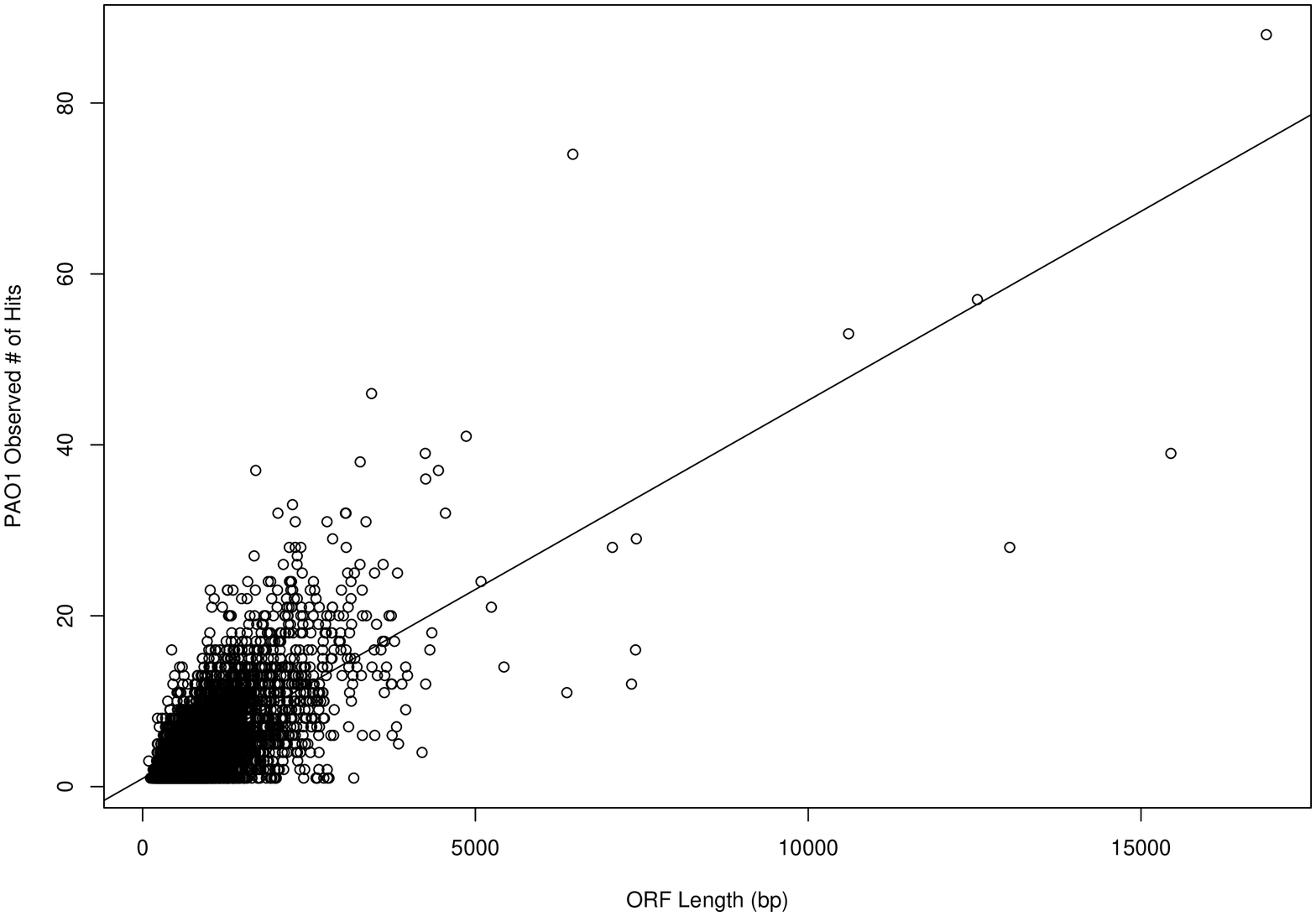}}}}

(b) \rotatebox{270}{\resizebox{2.8in}{4.0in}{{\includegraphics{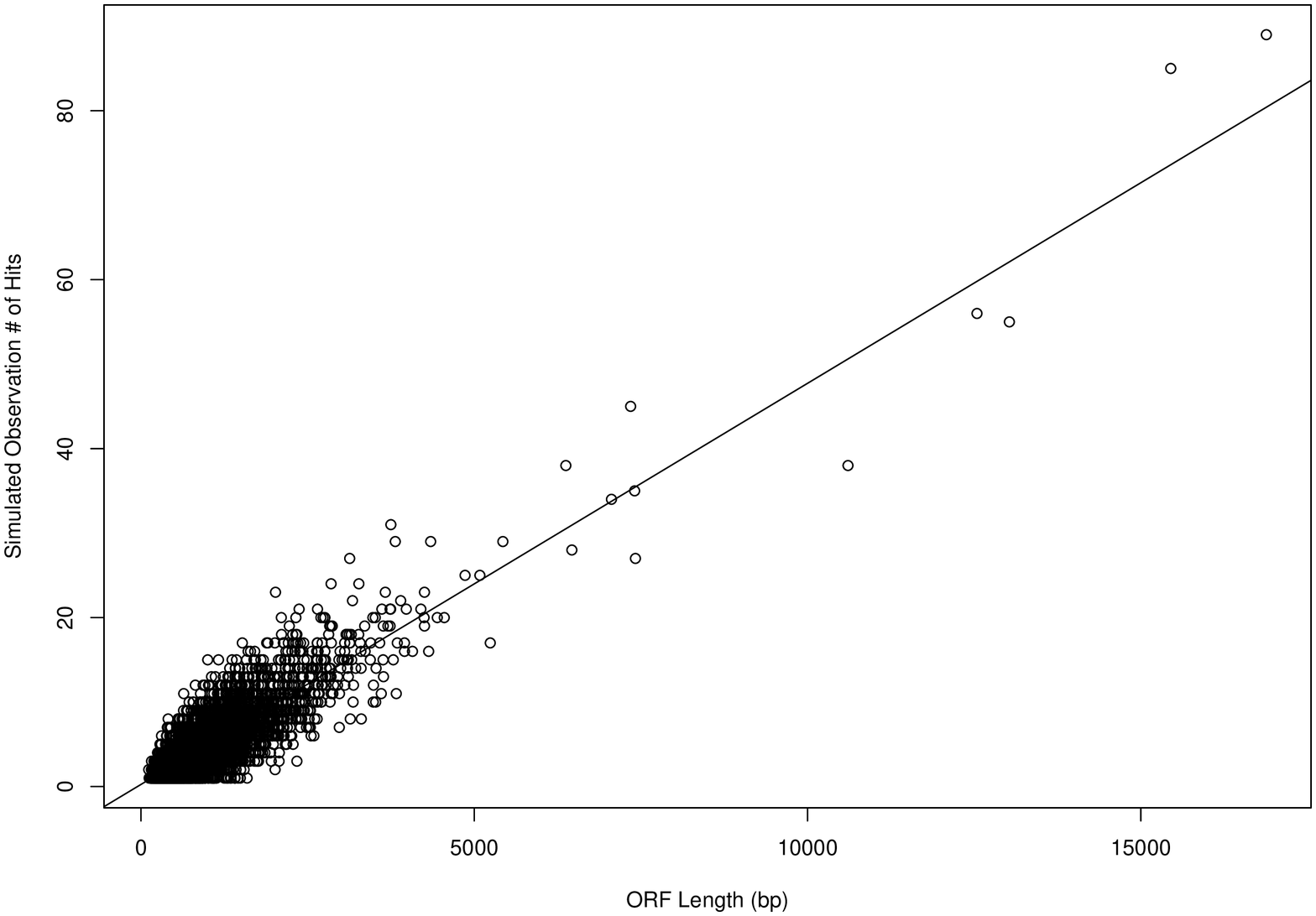}}}}
\caption{Comparison of the observed and theoretical probabilities
of insertion. Plot (a) is the observed number of insertions on ORF
length. (b) is a simulated number of insertions on ORF length.
Both plots only include ORFs that were hit at least
once.}\label{scatter}
\end{figure}

To observe the discrepancy between the predicted probabilities and
the experimental distribution of hits, we plot the observed number
of insertions on gene length in bp for genes that have been hit
once next to a simulated set of insertions from the correct
multinomial (Figure~\ref{scatter}). We see that the coefficient of
determination, $R^2$, for the actual observations is 0.45 and for
the simulated insertions it is 0.72 meaning that the transposons
inserted in some ORFs more often than we expect and into others
less. There is no evidence that this phenomenon corresponds with
ORF length because the slopes of the linear regressions for both
graphs are tantalizingly close: $4.42\times10^{-3}$ for the
observed data and $4.75\times10^{-3}$ for the simulated data.

For \mt data set, the number of TA dinucleotides in an
ORF divided by the number of TAs in the genome might accurately
predict the probability that the transposon inserts
into a gene, but our test does not have enough power to detect
model failure at the 0.3X coverage level. Using the list of
\mt insertions, we find a $\chi^{2*}$ fit of 854.8 which is not in
the 95\% rejection region for 4,279 degrees of freedom. This lack of power is due to the
low coverage level, not transposon or organismal variation. If we
take a sample of 1,839 PAO1 insertions, we find a fit
of 1288.1 on 6324 degrees of freedom.

\section{Comparison of Estimators}
Although the probabilities based on the length of the gene are
approximate, we can still compare the two methods for estimating
the number of essential genes. Here's a brief description of the parametric bootstrap.\cite{davison:bootstrap97} One can fit the function
\beq\label{owmj} f(n) = b_0-b_1\exp(-b_2n) \eeq

\no to the cumulative plot of the number of ORFs hit. The parameters  $b_0$, $b_1$, and $b_2$
are chosen to minimize the residual standard error between the
function and the data. Fitting 30,100 points is computationally
intensive, and we want to use exclusively R for all our
programming, so we choose 100 equally spaced points including the
last one to fit. Hence, for all the insertions, look at the number
of different ORFs hit at $301,602,\ldots,30100$ insertions. One
interprets the parameter $b_0$ as the number of nonessential
genes.

However, fitting this model is not a standard nonlinear regression,
so we compute the bias and variance of the estimated
parameters in a different manner. We proceed by assuming that
parameters fitted to the actual cumulative plot have the same bias
and variance as when they are fitted to the multinomial model
without any essential genes. In other words, $k-b_0^* \sim
m-\hat{m}$ where $\hat{m}$ is the estimate of the number of
nonessential genes. We simulate $(X,Y)$ according to
Equation~(\ref{dist}) with all $g_i=1$  by drawing a sample without
replacement of size $n$ from the set of targets and placing them in the appropriate ORF. If
$b_{0j}^*$ is the $b_0$ fitted to the $j^{\mbox{th}}$ simulated
experiment of $l$ experiments, $\mbox{Bias } \hat{m} \approx
\bar{b_0^*}-k$ where $k$ is the number of ORFs in the genome and
$\Var(\hat{m}) \approx \frac{1}{l-1}\sum_{1\le j \le
l}(b_{0j}^*-\bar{b_0^*})^2$. To compute the $100(1-\alpha)\%$
confidence interval, we use
\begin{eqnarray*}
P\left(\hat{m}+k-b^*_{0(1-\alpha/2)} \le m \le
\hat{m}+k-b^*_{0(\alpha/2)}\right) & \approx & 1-\alpha
\end{eqnarray*}

\no where $b^*_{0(x)}$ is the $x^{\mbox{th}}$ quantile.

We estimate that there are 382 essential
genes with a standard deviation of 15.8. The 95\% confidence
interval for the number of essential genes is [340,412]. 

We
find a posterior mean for the number of essential genes of 408
with a 95\% credible interval of [384,432] using the Gibbs sampler.  At a coverage of 5.4X insertions
per ORF, both estimates of the number of essential genes were
basically the same and agree with a basic bioinformatic
assessment of gene essentiality.\cite{jacobs:pa03}

\begin{figure}[ht]

\resizebox{4.7in}{2.9in}{\includegraphics{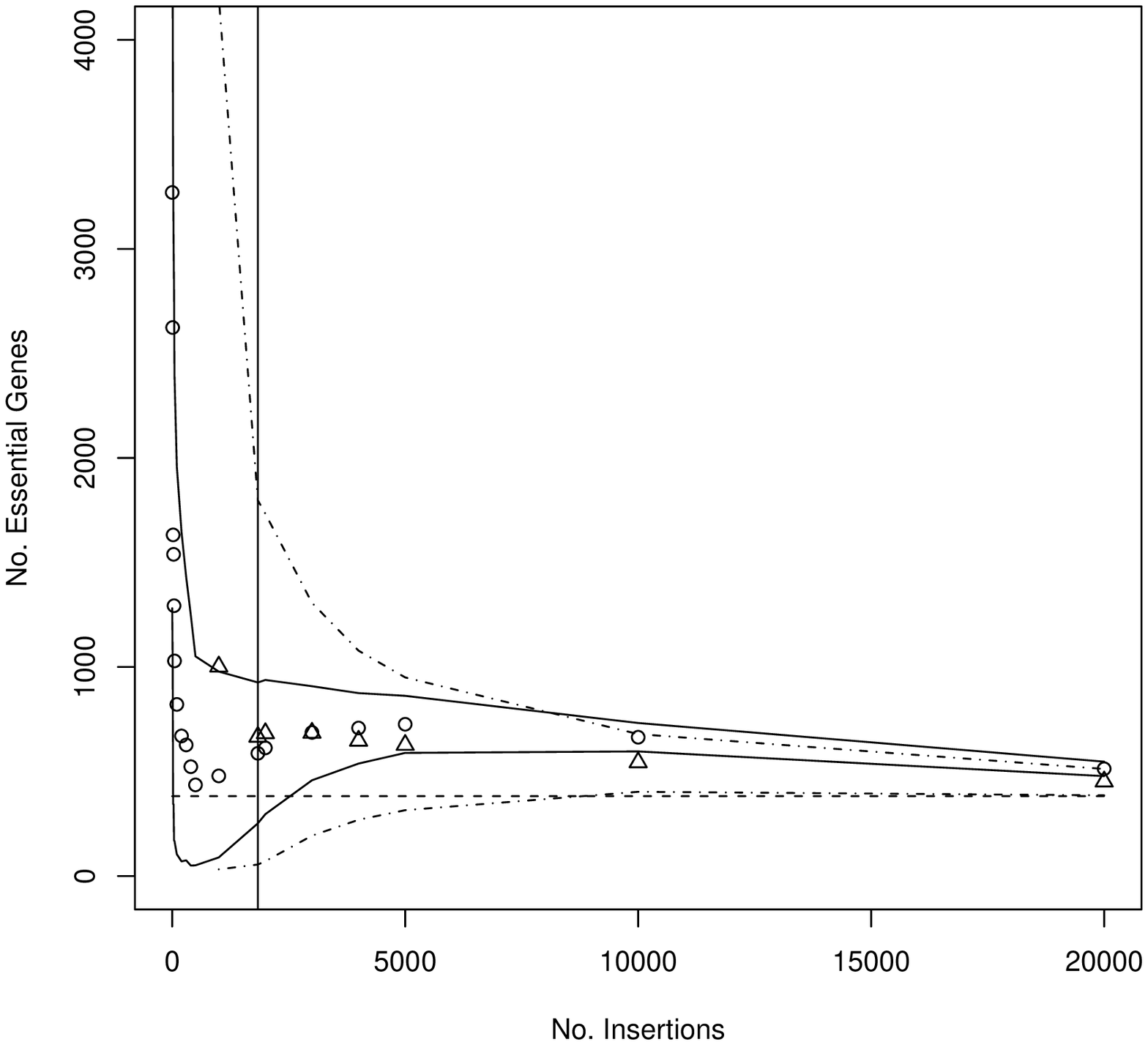}}
\caption{Mean point estimates and credible intervals for 100
simulated experiments based on the empirical distribution using
the PAO1 insertion list. We draw samples without replacement of
various sizes. The dashed, horizontal line is at 382, our estimate
of the number of essential genes. The vertical line is the
sampling level of the Lamichhane \etal paper. The
triangles are the points estimates from the parametric bootstrap
method and the circles are point estimates from the Gibbs sampler.
The dotted lines form the bounds of the 95\% confidence interval
and the solid lines the bounds of the 95\% credible
interval.}\label{estComp}

\end{figure}

To see how well our the parametric bootstrap method compared with the Gibbs sampler, we conduct a study using the empirical
distribution based on the locations of the 30,100 unique insertion
locations in the PAO1 clonal library. We assume that genes were
nonessential when an insertion landed anywhere in the entire
length of the ORF. We draw 100 samples of a certain coverage
without replacement from the 30,100 unique insertion locations and
for each sample, we compute the point and interval estimates
using both estimation procedures. The means of the 100 samples are
displayed in Figure~\ref{estComp}.

The Gibbs sampler has
tighter credible intervals overall but has more bias than the parametric bootstrap.  The
credible region at the 20,000 insertions (3.6X coverage) did not cover the estimate of the number of
essential genes at 30,100 insertions whereas the 95\% confidence
interval always did. The Gibbs sampler estimate also exhibits
some rather strange behavior at extremely low coverage levels. The
most accurate point estimate occurred at 500 insertions or 0.09X
coverage. We do not
see the immediate drop in the estimate as in
Figure~\ref{estComp} when we explore the behavior of the Gibbs sampler by
simulating insertions from a simplified, correct mode. Our results are not shown but they are
consistent with the simulation studies in Blades and Broman's technical report.\cite{blades:essOrf02} The Gibbs sampler on simulated data behaves
like the parametric bootstrap does on the experimental data. Therefore, we have little
faith in the accuracy of their method at coverage levels at around 0.1X .

We use the empirical distribution based on the 30,100 unique
insertions to find the number of clones needed for a 0.90 probability that the multinomial model would be rejected using the modified $\chi^{2*}$ statistics and for the 95\% credible
interval found by the Gibbs sampler would cover 408; their best
estimate of the number of essential genes. We draw a number of
samples without replacement of different coverage levels and
determine when the event happened about 90\% of the time. We determine that one needs a 2.9X coverage for
a 0.9 probability of the $\chi^{2*}$ test rejecting the null
hypothesis that the probabilities based gene length are correct. Secondly,
one needs a coverage of 5.0X to have a 0.9 probability of covering
408 essential genes using the Gibbs sampler's 95\% credible
interval.

\section{Discussion}
Numerous statistical questions arise while creating random
transposon mutagenesis clonal libraries. Two groups of
researchers developed different
methods to estimate the number of essential genes in a prokaryotic
organism given the genomic reference sequence and the number and
types of clones in the library.

First, both methods are based on a multinomial insertion model
that has a substantial number of parameters that must be computed
\textit{a priori}. For each ORF, we need to be able to calculate
 the probability that a transposon will land in the ORF given that it is nonessential based on the
reference sequence. For the
transposons used in the PAO1 library, one assumes that
each inserts at a nonessential bp of the genome with equal
probability. Likewise for the transposon used to build the \mt library, one
assumes that it inserts at each dinucleotide TA with equal
probability. It definitely inserts at only TA
dinucleotides,\cite{lampe:himar196} and Lamichhane \etal present
evidence that it inserts at each nonessential TA with equal
probability. We have duplicated their analysis for model fit. The PAO1 library transposons have little
preference for where they insert in a nonessential genes
(Figure~\ref{histOrf}). 9.4\% of the unique insertions land in
the 10.6\% of the PAO1 genome in intergenic regions whereas 17\%
of the insertions land in the 13\% of the \mt ~genome in
intergenic regions. The PAO1 library transposons hit duplicate sites more often
than one would expect by chance. The majority of identical hits
are from experiments run on the same day and many are
from the same plate. This strongly suggests that siblings (mutant
strains that divide prior to being plated) and
cross-contamination (which may occur at multiple stages from
picking colonies through PCR) create the artifact in the data,
rather than genuine patterns of exact duplication. Lamichhane's
methodology did not create as many artifacts. Our extreme
high-throughput method would be predicted
to create a higher proportion of artifacts, most of which would
look like exact duplicates.

Extending the model-fit analysis, we
compare the observed number of hits with the expected number hits
in ORFs that were hit at least once. We reject the model that
the probabilities of being hit are determined by gene length. The lengths are predictive of the correct
distribution (See Figure~\ref{scatter}) but not as accurate as the
empirical probabilities computed from a simulation of a
multinomial. We see this by comparing the $R^2=0.44$ for the
actual data with the $R^2=0.77$ of the simulated data. 

We feel that each gene confers a specific fitness to the cell
which causes the difference in coefficients of determination (the
$R^2$ values). Some are essential, so cells lacking these genes
die. Some are somewhat essential, so cells missing these genes can
live but not very well. Finally, some are completely nonessential,
so it does not matter whether cells have these genes. The idea
that cells lacking different functional genes have different
fitness is not novel.\cite{thatcher:marginal98,yu:marginal04} How
one would estimate these fitnesses is still being studied,
especially how to do it from reference sequence alone. Our guess
is that it would involve a competitive approach such as the one
taken by Gerdes \etal.\cite{gerdes:ess03}

Second, we compare how well the parametric bootstrap and the Gibbs
sampler do at estimating the number of essential genes at lower
coverage levels. The parametric bootstrap has less bias but a greater
variance (Figure~\ref{estComp}). We did a careful exploration
of model fit to understand why the Gibbs sampler appears to be
hyper-efficient. According to Figure~\ref{estComp}, one could
analyze 500 insertions, a coverage of 0.09X insertions per ORF and
find an accurate estimate of the number of essential genes. We do
not see the hyper-efficiency in our simulated data sets or the
simulated data sets in Blades and Broman's technical report\cite{blades:essOrf02} leading us to believe that mis-specified
insertion probabilities are causing the behavior.

Third, we estimate coverage levels required to achieve various
goals. The most critical determination is the coverage one needs
to reject the multinomial insertion model. For PAO1, three unique
insertions per ORF are needed. If we are interested in accurate
estimation of number of essential genes at sub-saturation coverage levels of around 0.3X, we believe that it is
mandatory that one has an accurate probability model. However,
this creates a paradox. We cannot know if the probability model is
correct unless we achieve at least a modest level of coverage. Lamichhane \etal should have mapped 12,750 insertions
to check model sufficiency rather than 1,425 and this would have
undermined their goal of looking at a low sub-saturation coverage
level. It is possible that a more powerful statistical test will
help us identify model inaccuracy.  If the
transposon used to build the \mt library were shown to insert with the correct probabilities in
a higher coverage level study, it would become the transposon of
choice for low coverage studies.

However, what if the best model that we can find is only partially
correct? Right now, we do not know what covariates to use to
adjust the probabilities. We could try non-parametric techniques,
but these still would probably require high coverage levels. We
are beginning to look at probability models that do not
assume we know the probability of transposon insertion for an
ORF. Alternatively, we could put our effort into only high
coverage level studies because with a large number of inserts
mapped. We have evidence that both methods find a sensible
estimate for the number of essential genes. At the coverage level
of 5.4X, we find 382 essential genes using the parametric bootstrap which is
similar to the estimate of 408 essential genes using the Gibbs sampler.
Both estimates are similar to the count of 350 \pa's genes that
are homologous to essential genes in \textit{E. coli},\cite{jacobs:pa03} as well as matching proportions of essential
genes from other studies.

\section*{Acknowledgments}
We thank Mike Steel, Elizabeth Thompson, and Margee Will for their helpful
comments. We thank the Allan Wilson Centre for funding.

\bibliographystyle{plain}
\bibliography{clonal_lib.bib}

\end{document}